\documentclass[fleqn,usenatbib]{mnras}
\usepackage[T1]{fontenc}
\usepackage{lmodern}
\usepackage{ae,aecompl}

\usepackage{natbib}
\usepackage{epsfig}
\usepackage{subfigure}
\usepackage{float}
\usepackage{color}
\usepackage{amsfonts}
\usepackage{units}
\usepackage{multirow}
\def\jcap{JCAP}
\usepackage{graphicx}	% Including figure files
\usepackage{amsmath}	% Advanced maths commands
\usepackage{amssymb}	% Extra maths symbols

\title[Secondary-electron radiation from SNRs]{Secondary-electron radiation accompanying hadronic GeV-TeV gamma-rays from supernova remnants}

\author[Huang et al.]{
Yan Huang,$^{1,2}$\thanks{E-mail: hyan623@pku.edu.cn}
Zhuo Li,$^{1,2}$
Wei Wang,$^{3}$
and Xiaohong Zhao$^{4,5,6}$
\\
$^{1}$ Department of Astronomy, School of Physics, Peking University, Beijing 100871, China; hyan623@pku.edu.cn\\
$^{2}$ Kavli Institute for Astronomy and Astrophysics, Peking University, Beijing 100871, China \\
$^{3}$ School of Physics and Technology, Wuhan University, Wuhan 430072, China\\
$^{4}$ Yunnan Observatories, Chinese Academy of Sciences, Kunming 650216, China\\
$^{5}$ Center for Astronomical Mega-Science, Chinese Academy of Sciences, Beijing 100012, China\\
$^{6}$ Key Laboratory for the Structure and Evolution of Celestial Objects, Chinese Academy of Sciences, Kunming 650216, China
}

\date{Accepted XXX. Received YYY; in original form ZZZ}

\pubyear{2019}

\begin{document}
\label{firstpage}
\pagerange{\pageref{firstpage}--\pageref{lastpage}}
\maketitle

\begin{abstract}
The synchrotron radiation from secondary electrons and positrons (SEPs) generated by hadronic interactions in the shock of supernova remnant (SNR) could be a distinct evidence of cosmic ray (CR) production in SNR shocks. Here we provide a method where the observed gamma-ray flux from SNRs, created by pion decays, is directly used to derive the SEP distribution and hence the synchrotron spectrum. We apply the method to three gamma-ray bright SNRs. In the young SNR RX J1713.7-3946, if the observed GeV-TeV gamma-rays are of hadronic origin and the magnetic field in the SNR shock is $B\ga0.5$mG, the SEPs may produce a spectral bump at $10^{-5}-10^{-2}$eV, exceeding the predicted synchrotron component of the leptonic model, and a soft spectral tail at $\ga 100$keV, distinct from the hard spectral slope in the leptonic model. In the middle-aged SNRs IC443 and W44, if the observed gamma-rays are of hadronic origin, the SEP synchrotron radiation with $B\sim 400 - 500 \mu$G can well account for the observed radio flux and spectral slopes, supporting the hadronic origin of gamma-rays. Future microwave to far-infrared and hard X-ray (>100keV) observations are encouraged to constraining the SEP radiation and the gamma-ray origin in SNRs.
\end{abstract}


%%%%%%%%%%%%%%%%%%%%%%%%%%%%%%%%%%%%%%%%%%%%%%%%%%

%%%%%%%%%%%%%%%%% BODY OF PAPER %%%%%%%%%%%%%%%%%%
\section{Introduction}
Supernova remnants (SNRs) are long considered to be the prime sources of Galactic cosmic rays (GCRs) for energies at least up to the CR spectral ``knee'' at $\sim 3 \times 10^{15}$eV, and probably up to the spectral ``ankle" at $\sim 10^{18}$eV, above which the CRs might come from extragalactic sources \citep{aha13, gab09}. There are mainly two reasons to believe that SNRs are the major sites to produce GCRs. The first is based on the produced rate of GCRs, $\dot W_{\rm CR} \approx \rm (0.3 -1) \times 10^{41} erg/s$, which can be supplied by approximately $10 \%$ of the kinetic energy of Galactic supernovae (SNe) \citep{gai91}. The second is that the diffusive shock acceleration (DSA) process in SNRs is expected to convert the kinetic energy of the bulk motion of SN ejecta to relativistic electrons and nuclei effectively. Many theoretic and observational works try to give more direct evidence to prove or disprove SNRs as GCR sources.

In the past decades, multi-wavelength observations had substantially increased our knowledge of SNR phenomena. Young and middle-aged SNRs are often observed to emit nonthermal radiation from radio up to GeV-TeV gamma-ray bands, which is believed to be produced by high energy electrons and/or protons accelerated by SNR shocks. The radio to X-ray emission can be originated from synchrotron radiation from the primary shock-accelerated electrons. However, the origin of high energy gamma-ray emission, whether Inverse Compton (IC) scattering/bremsstrahlung emission from primary electrons (leptonic model) or pion production caused by the collisions of accelerated protons/ions with background plasma (hadronic model), is less understood, because both leptonic and hadronic models can well explain the high energy GeV-TeV gamma-ray emission.

Some methods must be developed to distinguish the main production mechanism for gamma-ray emission from SNRs. At GeV energies two middle-aged SNRs, IC 443 and W44, show the characteristic spectral feature (often referred to as the ``pion-decay bump") uniquely identifies $\pi^{0}$-decay gamma rays and thereby high-energy proton existence \citep{ack13}. More direct evidence of proton-proton ($pp$) interactions should be detection of gamma-ray emission exceeding $\sim50$TeV, since the leptonic process suffers from Klein-Nishina suppression at such high energies. Future deep spectroscopic and morphological studies of SNRs with Cherenkov Telescope Array (CTA) and LHAASO promise a breakthrough regarding the identification of radiation mechanism \citep[e.g.,][]{aha13}.

Here, we present another potential method to distinguish the hadronic and leptonic models. In the hadronic process, the $pp$ interactions produce not only pionic gamma-rays but also secondary electrons and positrons (SEPs) from charged pion decays. The SEPs may compete with primary accelerated electrons and significantly contribute to the nonthermal electromagnetic radiation \citep[e.g.,][]{GAC09, pet16}. The SEP synchrotron radiation may differ from that contributed by the primary accelerated electrons. So the detection of the synchrotron radiation by SEPs can be an alternative method to distinguish the mechanisms of gamma-ray emission in SNRs.

Both SEPs and secondary gamma-rays are produced in $pp$ interactions, and their spectra both depend on the unknown CR spectrum and the medium density surrounding the SNR. However, there is a tight correlation between the SEPs' and secondary gamma-rays' spectra due to particle physics, so we can derive SEP spectrum by gamma-ray spectrum, and vise verse, avoiding knowing the uncertain CR and medium properties.
Here we provide in \S 2 a new method of deriving the SEP distribution and hence their synchrotron radiation directly by the observed gamma-ray spectrum and flux from SNRs.
%we gave a simple analytical method to transform the gamma-rays spectra to the SEPs spectra. Secondary electrons have been accumulated in the remnant, in order to determine the synchrotron %radiation of the SEPs, we must solve a simplified time-dependent electron transport equation.
In \S 3 we apply our method to several SNRs with GeV-TeV detections, RX J1713.7-3946, IC 443 and W44. \S 4 will be the discussion on the results, and the main conclusion is summarized in \S 5.

\section{Method}
The SN ejecta drives a shock into the surrounding medium. The shock swept-up particles can be accelerated by the SNR shock. The acceleration makes SNRs potential sources of GeV-TeV gamma rays, resulted from decays of secondary $\pi^{0}$-mesons produced in hadronic interactions.

A direct signature of high energy protons is provided by gamma rays generated in the decay of neutral pions ($\pi ^{0}$): $pp$ collisions create $\pi$ mesons, including neutral and charged $\pi$-mesons. The neutral pions will quickly decay into two gamma rays, each having an energy of $m_{\pi ^{0}} c^2/2=67.5$ MeV in the rest frame of the neutral pion. On the other hand, the charged pions will quickly decay into SEPs, as well as neutrinos. The decay channels are as following,
\begin{equation}
pp~\to
\begin{cases}
\pi^{0}+X ~ \to \gamma \gamma, \\
\pi^{+}+X ~ \to e^{+}  \nu_{e}  \nu_{\mu} \bar \nu_{\mu},\\
\pi^{-}+X~ \to e^{-}  \nu_{e}  \bar \nu_{\mu} \nu_{\mu}, \\
\end{cases}
\end{equation}
where $X$ represents other products. % such as $\eta$ meson, etc.
Here, we take the approximate isospin-invariant distribution of pions, i.e., the three types of pions produced have similar numbers at given energies,  $\pi^{+}:\pi^{-}:\pi^{0}\approx1:1:1$.
Thus there will be a correlation between the energy distributions of SEPs and gamma-ray photons. In the following we provide a way to  derive the distribution of the SEPs directly by using the gamma-ray emission. In the derivation we adopt the parameterized distribution functions of secondaries in $pp$ interactions from \citet{kel06}. At last we calculate the synchrotron radiation by the generated SEPs in the SNR shock.

\subsection{SEP energy spectra from $pp$ interactions}
\label{sec:SEP}
Define $J_{\pi} (E_{\pi})$ as the spectrum of $\pi^{0}$-mesons produced in $pp$ interactions, i.e., the produced number per unit pion energy $E_\pi$ per unit time $t$, then the produced rate of pions in the energy interval ($E_{\pi}$, $E_{\pi}+dE_{\pi}$) is $d\dot{N}_{\pi}\equiv J_{\pi} (E_{\pi}) dE_{\pi}$. The energy distribution of gamma-rays from the decay of $\pi^{0}$-mesons, $\pi^{0} \to \gamma \gamma$, $Q_{\gamma}(E_{\gamma})\equiv d\dot{N}_{\gamma}/dE_{\gamma}$, is
\begin{align}
Q_{\gamma}(E_{\gamma})=2 \int_{E_{\gamma}}^{\infty} J_{\pi} (E_{\pi}) \frac{dE_{\pi}}{E_{\pi}}.
\label{gamma}
\end{align}
On the contrary we may obtain the energy distribution of gamma-ray production rate by observed photon flux ($\Phi$: photon number per unit gamma-ray energy per unit detector area per unit time) $Q_\gamma=4\pi D^2\Phi$, with $D$ the source distance from the Earth. Given $Q_{\gamma}(E_{\gamma})$, Eq.(\ref{gamma}) can be used to derive the $\pi^{0}$-mesons distribution $J_{\pi} (E_{\pi})$,
\begin{align}
J_{\pi} (E_{\pi})=-\frac{1}{2} \left[E_{\gamma}\frac{dQ_{\gamma}(E_\gamma)}{dE_{\gamma}}\right]_{E_{\gamma} \to E_{\pi}}.
\label{pi}
\end{align}
Eq.(\ref{pi}) shows that the $\pi^{0}$-mesons distribution depends on the derivative of the gamma-ray spectrum, $Q_{\gamma}(E_{\gamma})$.
%(for $E_{\gamma}$) first, then multiply by $E_{\gamma}$, next change $E_{\gamma}$ to $E_{\pi}$ for the result of the first two steps. Final, multiply by a constant, $-1/2$.
The energy production rate of gamma-rays is equal to that of pions,
%\begin{align}
$\int_{0}^{\infty} E_{\gamma} Q_{\gamma}(E_{\gamma}) dE_{\gamma}= \int_{0}^{\infty}E_{\pi} J_{\pi} (E_{\pi}) dE_{\pi}$.
%\end{align}

%\subsubsection{$\pi^{\pm} \to e+\nu_{e}+2\nu_{\mu}$}\

The produced electrons are the secondary products of decays of charged pions, $\pi^{\pm} \to e+\nu_{e}+2\nu_{\mu}$. The spectrum of the produced SEPs, $Q_e(E_e)\equiv d\dot{N}_e/dE_e$, is given by
\begin{align}
Q_{e}(E_{e})=2 \int_{0}^{1} f_{e}(x) J_{\pi} (E_{e}/x) \frac{dx}{x},
\label{electron}
\end{align}
where $x=E_{e}/E_{\pi}$, and the factor 2 takes into account the contributions of both $\pi^{+}$ and $\pi^{-}$. The function $f_e(x)$ is given by:
\begin{align}
f_{e} (x) = g_{\nu_{\mu}}(x) \Theta(x-r) + (h^{(1)}_{\nu_{\mu}}(x) + (h^{(2)}_{\nu_{\mu}}(x))  \Theta(r-x),
\end{align}
where $r=(m_{\mu}/m_{\pi})^2=0.573$,
\begin{align}
g_{\nu_{\mu}}(x)= \frac{3-2r}{9(1-r)^2} (9x^2-6 \ln x -4x^3 -5),
\end{align}

\begin{align}
h^{(1)}_{\nu_{\mu}}(x)= \frac{3-2r}{9(1-r)^2} (9r^2-6 \ln r -4r^3 -5),
\end{align}

\begin{align}
h^{(2)}_{\nu_{\mu}}(x)= \frac{(1+2r)(r-x)}{9r^2} [9(r+x)-4(r^2+rx+x^2)],
\end{align}
$\Theta$ is the Heaviside function ($\Theta(x)=1$ if $x \geq 0$, and $\Theta(x)=0$ otherwise). Thus, given the pion-decayed gamma-ray spectrum $Q_\gamma(E_\gamma)$, one can use Eq.(\ref{pi}) and Eq.(\ref{electron}) to derive the energy distribution of produced SEPs, $Q_e(E_e)$.

Some notes should be made here on the method. First, eq. (\ref{gamma}) (so that eq. \ref{pi}) is valid for $E_{\pi} \gg m_{\pi} c^2\approx140$~MeV, whereas for $E_{\pi} \sim m_{\pi} c^2$, the precise equation is eq. (78) in \citet{kel06} (also see \citet{kaf14}). However, it can be found that our method to derive SEP spectrum using eqs. (\ref{pi}) and (\ref{electron}) is good enough for gamma-rays and SEPs with energy of $E_{\gamma}\ga100$~MeV and $E_e\ga100$~MeV, respectively. Second, we ignore the gamma-ray flux from $\eta$ meson decay, which only contributes about $10\%$ of the total flux from $pp$ interactions \citep{kel06}.

\subsection{Synchrotron radiation from SEPs}
%Once the SEPs are produced in the SNR, they will emit their energy and cool mainly by synchrotron radiation and/or IC scattering background photons.
The SEPs are accumulated in the SNR, but their energy distribution is affected by radiative cooling. We derive first the energy distribution of SEPs, and then calculate the synchrotron spectrum emitted by them.

We solve the continuity equation governing the temporal SEPs' distribution
%\footnote{To explain the radio flux in this work, one zone model is enough, so we simply take a one zone model here.},
\begin{align}
\label{con}
\frac{\partial}{\partial t} \frac{dN_{e}}{d\gamma_{e}} +\frac{\partial}{\partial \gamma_{e}} \left[\dot \gamma_{e} \frac{dN_{e}}{d\gamma_{e}}  \right] = S(\gamma_{e},t).
\end{align}
Here $\gamma_{e}\equiv E_e/m_ec^2$ is the Lorentz factor of the SEPs, $dN_{e}/d\gamma_{e}$ is the instantaneous SEPs' spectrum in the SNR at time $t$ ($t=0$ denotes the time of the SN explosion), and $S(\gamma_{e},t)\equiv Q_e(E_e,t)m_ec^2$ is the source function.
%{\bf The production rate of secondary electrons, $Q_e(E_e,t)$, to be constant since that most of particles have been accelerated and lost their energy in the downsteam of the postshock in a dynamical timescale.}
We can obtain $S(\gamma_e,t)$, e.g., by gamma-ray observations, by the way explained in \S \ref{sec:SEP}.

\begin{figure}
\vskip -0.0 true cm
\centering
\includegraphics[scale=0.4]{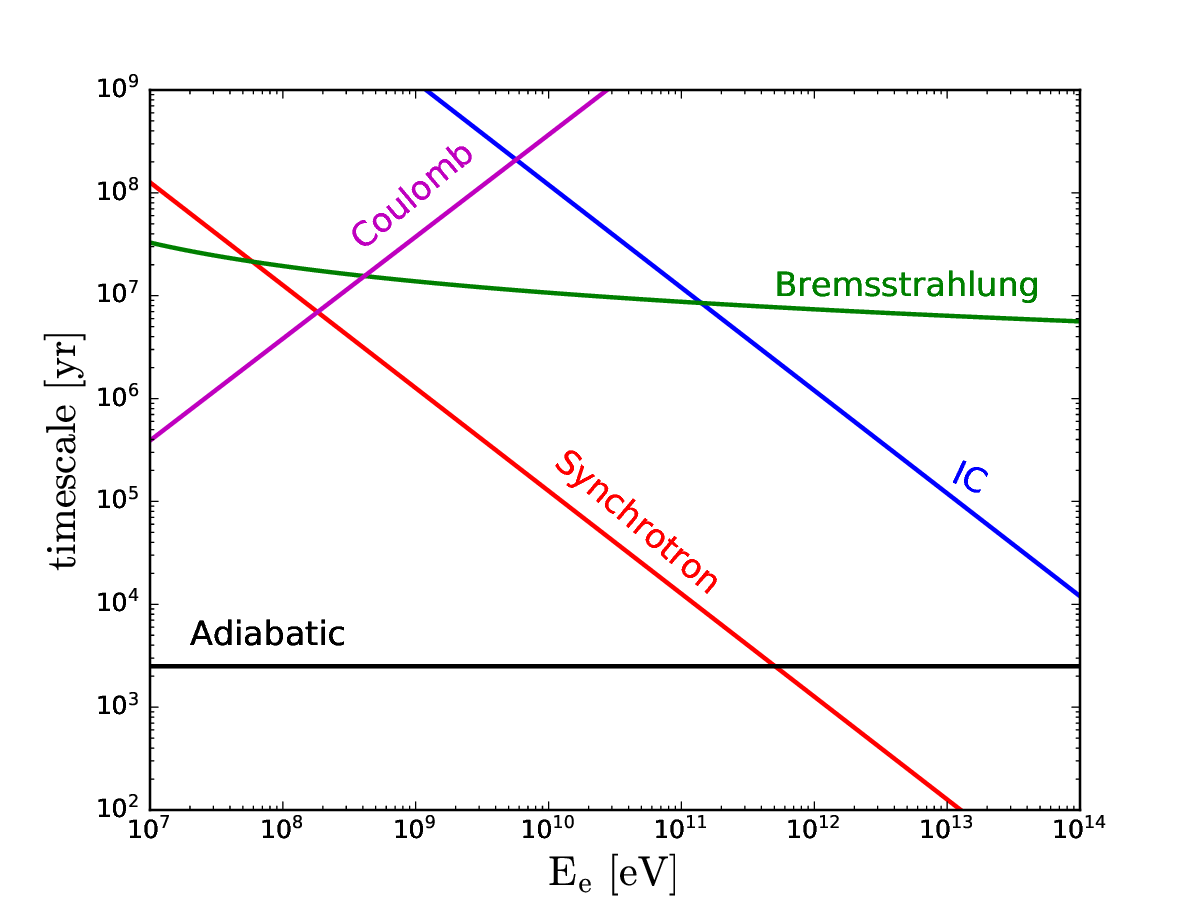}
\includegraphics[scale=0.4]{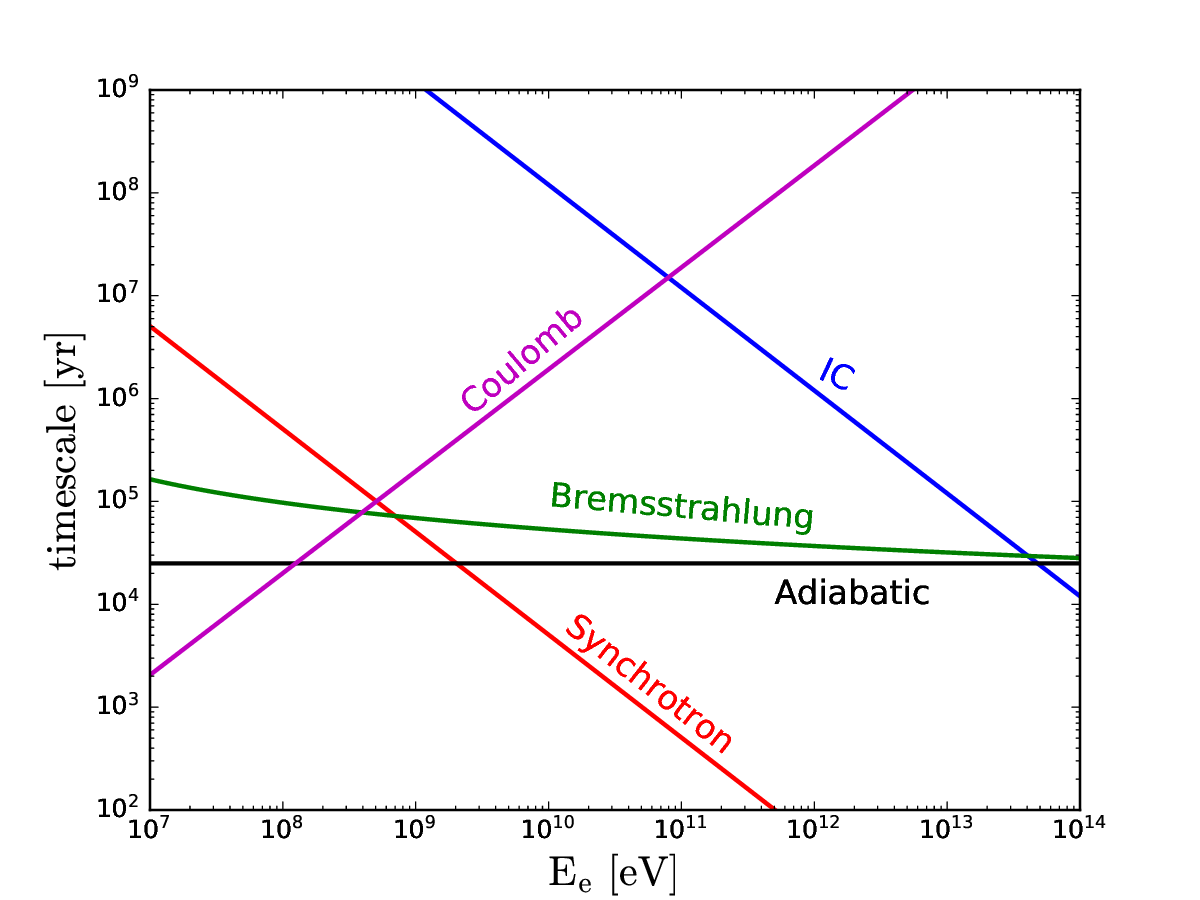}
\caption{Energy-loss timescales as function of electron energy. Upper panel: $n_{\rm H}=n_e=\rm 1 cm^{-3}$, $B=0.1 \rm mG$ and $t=1 \rm kyr$. Lower panel: $n_{\rm H}=n_e=\rm 200 cm^{-3}$, $B=0.5 \rm mG$ and $t=10 \rm kyr$.}
\label{Fig:timescale}
\end{figure}

The energy loss rate of SEPs, $\dot\gamma_e$, is mainly determined by several cooling processes:
\begin{align}
\label{loss}
\dot \gamma_{e}=\dot \gamma_{e,\rm  coul} +\gamma_{e,\rm  brems}+\dot \gamma_{e,\rm  syn} + \dot \gamma_{e,\rm  IC} + \gamma_{e,\rm  ad}.
\end{align}
The first is Coulomb collisions between energetic electrons and background electrons \citep{rep79},
\begin{align}
\label{coul}
\dot \gamma_{e,\rm  coul}=-\frac{3}{2} c \sigma_T n_e \ln \Lambda(\gamma_e,n_e),
\end{align}
where $\sigma_T$ is the Thomson cross-section, $n_e$ is the background electron density, and $ \ln\Lambda(\gamma_e,n_e)\sim40$ is the Coulomb logarithm. The second is the bremsstrahlung energy loss. In fully ionized hydrogen plasma we have \citep{ski96}
\begin{align}
\label{brems}
\dot \gamma_{e,\rm  brems}=-2 \alpha c \sigma_{T} n_{\rm H} \gamma_{e} (\ln \gamma_{e} +0.36),
\end{align}
where $n_{\rm H}$ is the postshock hydrogen density and $\alpha$ is the fine structure constant. The third is synchrotron radiation, given by \citep{ryb79}
\begin{align}
\label{syn}
\dot \gamma_{e,\rm  syn} = -\frac{\sigma_{T} B^2 \gamma_{e} ^2} {6 \pi m_{e} c},
\end{align}
where $B$ is the postshock magnetic field strength. The forth is inverse Compton (IC) up-scattering the soft photons. The cosmic microwave background (CMB) photons dominate the soft background photons. In the Thomson regime, the energy-loss rate via IC scatterings of the CMB field is given by  \citep{ryb79}
\begin{align}
\label{IC}
\dot \gamma_{e,\rm  IC} =-\frac{4 \sigma_{T}  \gamma_{e} ^2 U_{\rm CMB}}{3 m_{e} c},
\end{align}
where $U_{\rm CMB} \approx 0.26 \rm eV\, cm^{-3}$ is the energy density of CMB photons. The last one is the adiabatic cooling due to the expansion of SNRs \citep{lon94}
\begin{align}
\label{adi}
\dot \gamma_{e,\rm  ad}=- \frac{\zeta \gamma_{e}}{t},
\end{align}
where $t$ is the time since SN explosions, and $\zeta$ is analytically derived to be $2/5$ in the SNR Sedov phase. In Fig.\ref{Fig:timescale}, we compare the energy-loss timescales of these processes, adopting the physical parameters that mimic the cases of SNRs RX J1713.7-3946 (upper panel) and IC 443 (or W44; lower panel). It can be seen that in the electron energy that we are interested in ($E_e\ga100$~MeV; see below), the dominant cooling processes are synchrotron and adiabatic cooling, and that IC cooling can be neglected.

%Given the low density in SNRs, the cooling of SEPs by bremsstrahlung radiation is neglected compared with synchrotron and/or IC cooling. As seen below we will only care about the evolution of the SNR system within one dynamical time, thus the adiabatic cooling of SEPs can also be ignored. Thus the cooling rate $\dot\gamma_e$ of an electron is given by

%\begin{align}
%\label{loss}
%\dot \gamma_{e}=-\frac{\sigma_{T} B^2 \gamma_{e} ^2} {6 \pi m_{e} c}(1+Y),
%\end{align}
%where $B$ is the magnetic field strength, and $Y$ is the Compton parameter parameterizing the IC cooling importance. In the following cases we consider SNR magnetic fields of $>10\mu$G, which has energy density much larger than that of the dominant photon background, i.e., the cosmic microwave background. Thus $Y\la U_{\rm ph}/U_{\rm B} \ll1 $, and we simply take $Y=0$ in Eq.(\ref{loss}).

We have neglected any escape of SEPs in Eq.(\ref{con}). This is because only the particles with energy around the maximum accelerated energy may be able to escape upstream of the shock, if the maximum energy is constrained by the finite dynamical time. The SEP energy is only a fraction of the primary cosmic rays, so they are confined and cannot escape from the system. Following \citet{cha70, chi99}, we use the fully implicit difference scheme to numerically solve Eq.(\ref{con}).

Given the solved-out distribution of SEPs, $dN_{e}/d\gamma_{e}$, the optically-thin synchrotron radiation produces a total luminosity per unit photon frequency $\nu$ of \citep{cru86}
\begin{equation}
L_\nu=\frac{\sqrt{3} e^3 B}{2 \pi m_{e} c^2} \int d\gamma_e \frac{dN_{e}}{d\gamma_{e}}  R(\nu/\nu_{c}),
\end{equation}
where $R(\nu/\nu_{c})$ describes the synchrotron radiative power of a single electron in magnetic filed with chaotic directions, for which we take a simple analytical form following \cite{zir07},
\begin{equation}
R(\nu/\nu_{c})=\frac{1.81\exp(-\nu/\nu_{c})}{\sqrt{(\nu/\nu_{c})^{-2/3}+(3.62/\pi)^2}},
\end{equation}
with $\nu_{c}=3eB \gamma_e^2/ 4\pi m_{e} c$ being the critical frequency for electrons with $\gamma_e$. At low enough frequencies, the correction due to synchrotron and free-free absorption may need to take into account.
The observed synchrotron radiation flux is then
\begin{equation}
F_{\nu}=L_\nu/4 \pi D^2.
\end{equation}

Note that our method is valid to derive SEPs of $E_e\ga100$~MeV, corresponding to Lorentz factors of $\gamma_e\ga10^2$. Depending on the magnetic field $B$ in the SNR shock, the related energy range of the synchrotron radiation is $h\nu\ga 10^{-7}(E_e/100 {\rm MeV})^2(B/1 {\rm mG}) \rm eV$. Thus, in the calculation we take a lower limit of the SEP Lorentz factor, $\gamma_{e,\min}\sim10^2$. This does not affect our discussion on the radio data at $\ga10^{-7}$eV, even the magnetic field is as high as $B\sim1$~mG.

We approximately consider a constant GeV-TeV gamma-ray luminosity, and hence a constant source function in Eq (\ref{con}) within the SNR age $t_{\rm SNR}$. We solve out the time-dependent $dN_e/d\gamma_e$ up to time $t=t_{\rm SNR}$, and then calculate the synchrotron spectrum of the same time.
The approximation of constant gamma-ray luminosity is reasonable as explained below. After the SN shock starts to decelerate and go into the adiabatic expansion phase, the total shock energy keeps a constant. If a constant fraction of the energy of the injected plasma goes to CRs, even though both the plasma and the CRs suffer adiabatic cooling downstream, the CR fraction keeps constant. Thus, the total CR energy in the shock is constant. The pion production rate, depending on total CR energy and medium density, is also constant.

The main uncertainty for the SEP synchrotron radiation is the magnetic field $B$ in the postshock region of the SNR shock, which we will regard as a free parameter. It is noted that in hadronic model for the GeV-TeV gamma-ray emission from SNRs a magnetic field of order of $100$'s $\mu$G is usually derived from spectral fitting.

\section{Application}

We will apply the method described in the previous section to derive the synchrotron radiation from SEPs directly with the observed gamma-ray spectrum, assumed to be hadronic origin.

\subsection{RX J1713.7-3946}

\begin{figure}
\vskip -0.0 true cm
\centering
\includegraphics[scale=0.4]{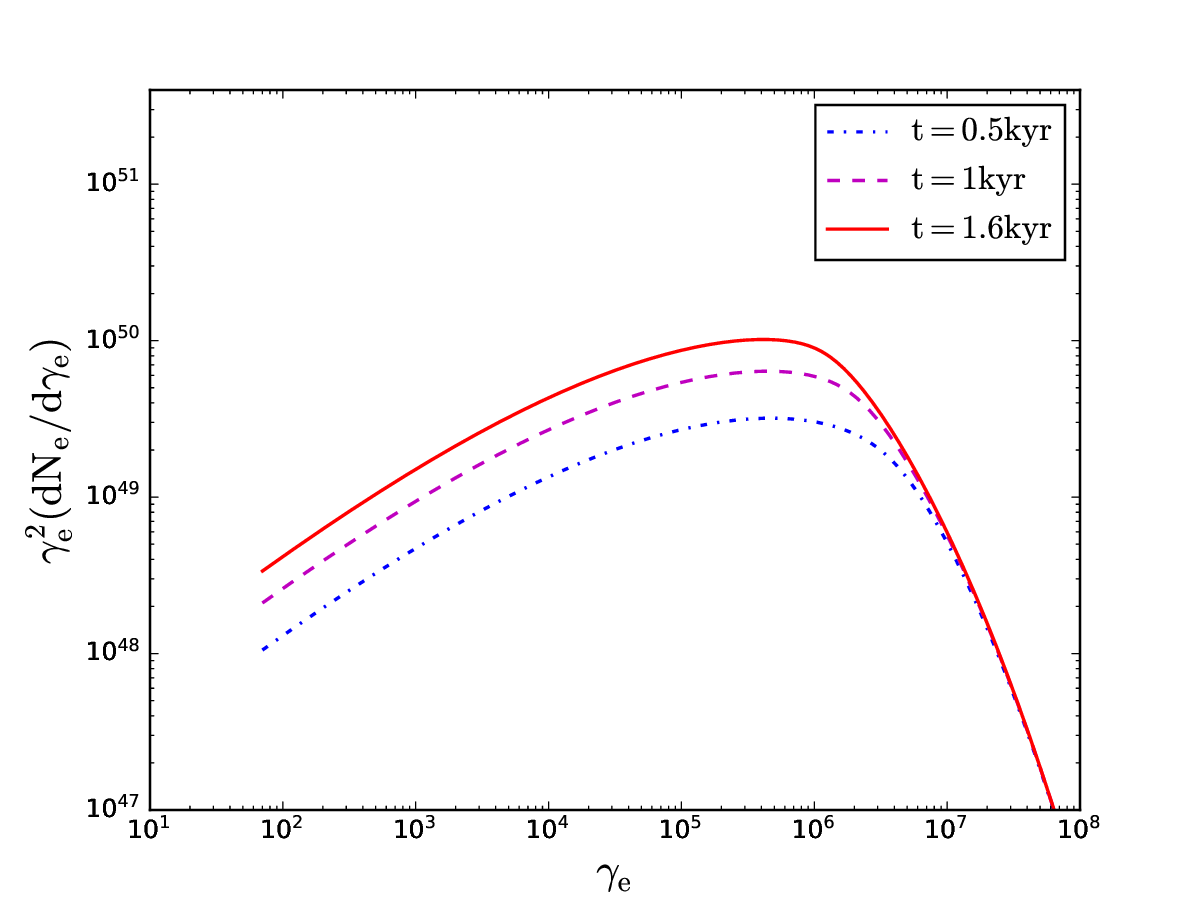}
\includegraphics[scale=0.4]{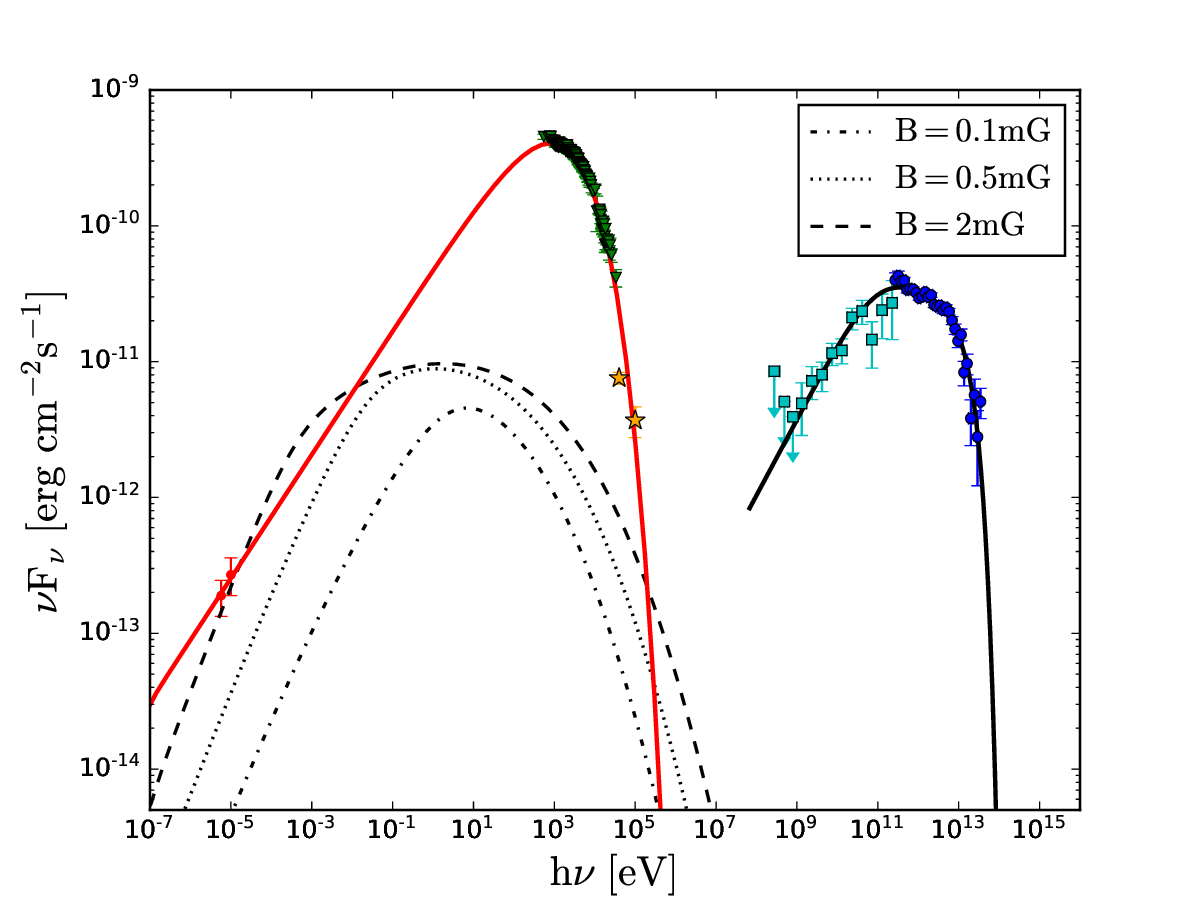}
\caption{Derived synchrotron radiation from SEPs in young SNR RX J1713.7-3946. Upper panel: temporal distribution of SEPs, assuming hadronic process dominates the production of GeV-TeV gamma-rays. The blue dashed-dotted, purple dashed, and red solid lines represent distributions at $t=\rm 0.5 kry$, $\rm 1 kyr$, and $\rm 1.6 kry$, respectively, with the postshock magnetic field $B=\rm 100 \mu G$. Lower panel: synchrotron spectrum from SEPs. The the black dashed, dotted, and dashed-dotted lines show the SEP synchrotron radiation at $t=\rm 1.6 kyr$ with different magnetic field $B=\rm 2 mG$, $\rm 0.5 mG$, and $\rm 0.1 mG$, respectively. The black solid line shows the Eq.(\ref{scpl}) spectral model to fit the gamma-ray data. The red solid line is the best fit of radio-to-X-ray data by synchrotron radiation of primary electrons with $B=0.5$ mG. Also shown are the radio data (red points) from ATCA \citep{laz04}, X-ray data (green triangles) from Suzaku \citep{tan08}, soft gamma-ray data (orange stars) from INTEGRAL-IBIS \citep{bir10}, $>0.1$ GeV data (cyan squares) from Feimi-LAT \citep{abd11}, and $>0.1$ TeV data (blue circulars) from HESS \citep{hes16}.}
\label{Fig:M1}
\end{figure}

\begin{table*}
	\centering
	\caption{Parameters used in gamma-ray spectral models.}
	\label{Tab:spectral fitting}
	\begin{tabular}{lcccccc}
		\hline
		SNR  & $K$ $[\rm 10^{-10} erg$ $\rm cm^{-2} s^{-1}]$ & $p_1$ & $p_2$ & $\xi$ & $E_1$ & $E_2$\\
		\hline
		RX J1713.7-3946  & $ 0.53$ &$ -0.56 $ &  $0.18$ & $1.38$ & $\rm 0.11 TeV$  &$\rm 18.34 TeV$\\
        IC 443   & $9.08 $ &  $0.74$ & $1.28$ & $-0.24$ & $\rm 8.67 GeV$ & $\rm 214.55 GeV$\\
        W44  & $17.45$  &  $1.60$ & $1.95$ &  $-0.34 $  & $\rm 4.25GeV$  & $ \rm 93.09 GeV$\\
		\hline
	\end{tabular}
\end{table*}

\begin{figure}
\vskip -0.0 true cm
\centering
\includegraphics[scale=0.4]{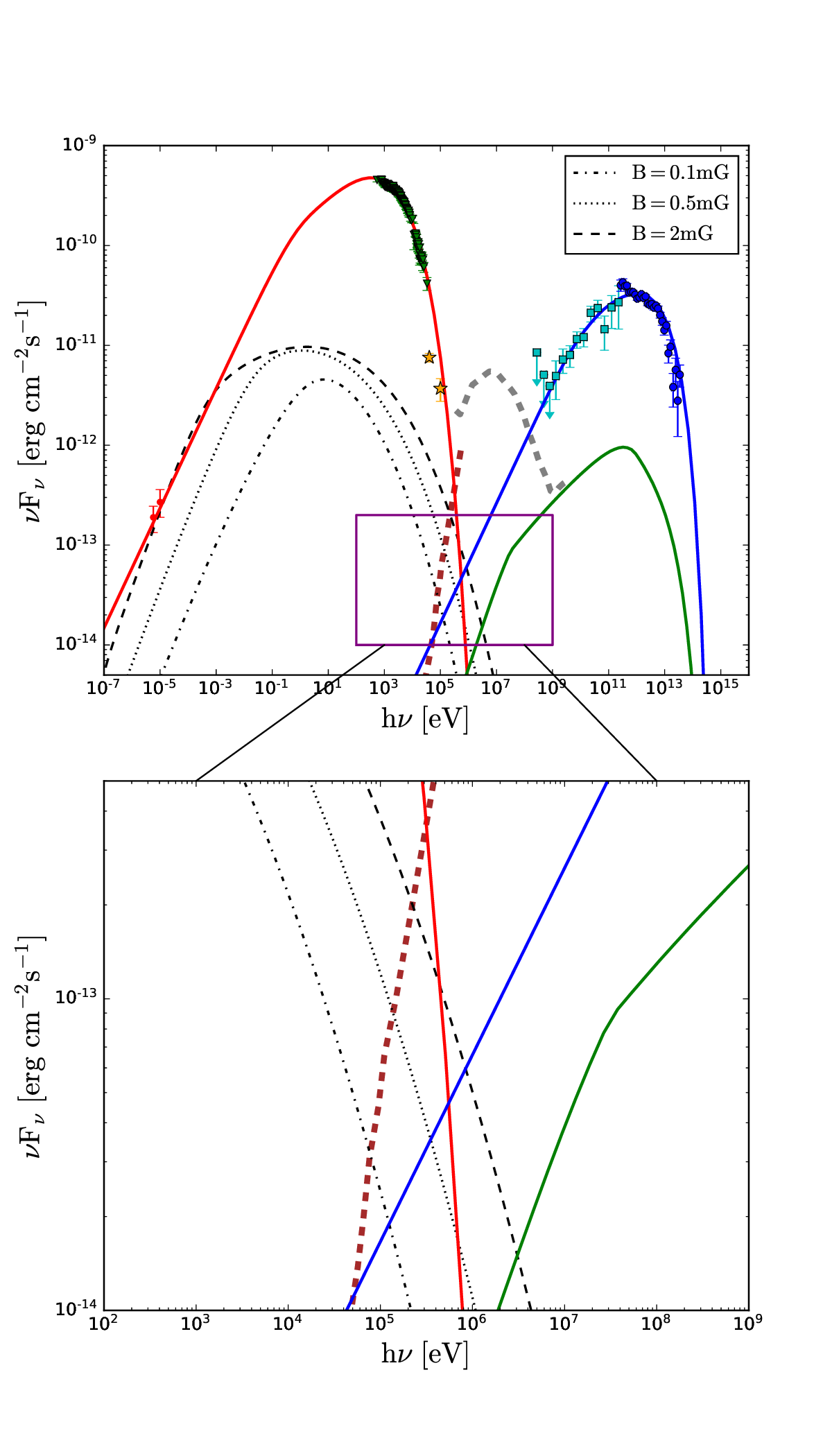}
\caption{Comparison of leptonic and hadronic models for the GeV-TeV emission in RX J1713.7-3946. The solid lines are the maximum likelihood leptonic model fit to the broadband data, using Naima fitting code; the red solid line is the synchrotron spectrum, the blue solid line is the IC radiation by upscattering the cosmic microwave background radiation and a far-infrared background radiation with temperature $T=26.5$ K and energy density $ 0.415\rm eV~cm^{-3}$\citep{hes16, shibata11}, and the green solid line is the corresponding nonthermal bremsstrahlung radiation for an ion number density $n= 1\rm cm^{-3}$. For comparison, the black dashed, dotted, and dashed-dotted lines are as same as in Fig.\ref{Fig:M1}, presenting the SEP radiation in the hadronic model, where the GeV-TeV emission is hadronic origin. The sensitivities of two future telescopes are shown: PHEMTO \citep{lau19} (brown thick dashed line) and e-ASTROGAM \citep{tat16} (grey thick dashed line).
}
\label{Fig:M2}
\end{figure}

RX J1713.7-3946 is the best studied young ($t_{\rm SNR} \approx 1.6 \unit{kyr}$) gamma-ray SNR. It was discovered in the $\mathnormal{ ROSTA}$ all-sky survey \citep{pfe96} and had an estimated distance of $D\approx 1$~kpc \citep{fuk03}. It is a prominent and well studied example of a class of X-ray bright and radio dim shell-type SNRs \citep{laz04}. Despite the past deep HESS exposure and detailed spectral and morphological studies \citep{hes16}, the origin of the gamma-ray emission (leptonic, hadronic, or a mix of both) is not clearly established\citep{mor09, zir10, ell10, yua11}.
For the leptonic model to explain the multiwavelength emission of RX J1713.7-3946, a small magnetic field ($B \approx \rm 14 \mu G$) \citep{hes16,tan08,yua11} is needed. However, for the hadronic model to explain the multiwavelength emission, a larger magnetic field ($B > 100 \mu G$) is needed \citep{zir10, yua11, gab14}. The observed rapid X-ray variability of RX J1713.7-3946 has been interpreted as a large, multi-mG, magnetic field \citep{uch07} \citep[see, however, e.g.,][]{katz08}.

Here, we apply our method to RX J1713.7-3946. We assume that the gamma-ray emission is due to the hadronic model. We fit the observed gamma-ray emission by a function as:
\begin{align}
E_{\gamma}^2 \Phi=K \left(\frac{E_{\gamma}}{E_{1}}\right)^{-p_1}\left[1+\frac{E_{\gamma}}{E_{1}}\right]^{p_1-p_2} \exp\left[-\left(\frac{E_{\gamma}}{E_{2}}\right)^{\xi}\right],
\label{scpl}
\end{align}
where $K$, $E_1$, $E_2$, $p_1$, $p_2$ and $\xi$ are parameters determined by fitting data. The best fitting parameters are showed in Tab.\ref{Tab:spectral fitting}. The parameter $\xi$ is positive, indicating that there is a spectral cutoff at high energies. We then use Eq.(\ref{pi}) and Eq.(\ref{electron}) to transform the spectra of observed gamma-ray to the spectra of SEPs, solve the time-dependent continuity equation and calculate the SEP synchrotron radiation. The negligible synchrotron and free-free absorption in RX J1713.7-3946 are ignored in the calculation. The results are showed in Fig.\ref{Fig:M1}. The upper panel in Fig.\ref{Fig:M1} represents the temporal electrons distribution which is the result governed by solving the continuity equation. The lower panel in Fig.\ref{Fig:M1} represents the synchrotron radiation of SEPs.

By changing the only free parameter $B$, we see in Fig.\ref{Fig:M1} that the synchrotron spectral profile of SEPs broaden as $B$ increases. For $B\simeq2$mG, the flux matches the observed radio flux by ATCA. So, if the gamma-rays from RX J1713.7-3946 are dominated by hadronic interactions, the magnetic field in the shock cannot be larger than $\sim$mG, otherwise the synchrotron radiation from SEPs overshoots the observed radio flux.

The radio-to-X-ray flux is usually believed to be dominated by the synchrotron radiation by primary electrons accelerated by the SNR shocks. We show in Fig. \ref{Fig:M1} a best fit of the radio-to-X-ray data by the synchrotron spectrum of primary electrons, using the public code Naima\footnote{https://github.com/zblz/naima}. The primary electrons' energy distribution is assumed to follow a broken power law with an high-energy exponential cutoff, and the postshock magnetic field is $B=0.5$ mG (Note the $B$ value is not determined because of the degeneracy of the electron distribution parameters and $B$). Comparing the radiation of primary electrons and SEPs in Fig. \ref{Fig:M1}, it can be found that for large magnetic field $B\ga0.5$mG the SEP radiation may exceed that of the primary electrons in the range of $10^{-5}-10^{-2}$eV. The flux level could be $\ga10^{-12}\rm erg\,cm^{-2}s^{-1}$. This is bright enough to be detected by many telescopes of relevant range. However, since that this source is located on the Galactic plane, and it is close to a HII region \citep{ace09}, removing the stronger contaminations is the main problem. At $\ga100$keV, the synchrotron radiation of SEPs may also show up as an excess compared to the synchrotron spectrum by primary electrons, but with a flux of only $\sim10^{-13}\rm erg\,cm^{-2}s^{-1}$ or below. Note that the spectra of IC and bremsstrahlung radiation by both the primary electrons and SEPs are all negligible in the case here, i.e., the magnetic field $B>0.1$mG and the density $n\sim1\rm cm^{-3}$ \citep{hes16}, thus they are not shown in the figure.

%\citet{abd11} shows that the spectrum of RX J1713.7-3946 in the Fermi-LAT energy range can be described by a hard power law with a photon index of $\Gamma=1.5$. In the leptonic model, the synchrotron spectral slope is the same as the IC one, thus the radio spectrum should be $\nu F_\nu\propto\nu^{0.5}$. We show such a spectral slope in the lower panel of Fig.\ref{Fig:M1} by fitting the radio data to present the prediction in leptonic model. The synchrotron radiation from SEPs in hadronic model can exceed the predicted flux in the leptonic model in $10^{-5}-10^{-2}$ eV if the magnetic field in the SNR shock is $\gtrsim \rm 0.5 mG$, and the observed flux could be above $\ga10^{-12}\rm erg\,cm^{-2}s^{-1}$.

We emphasize that the spectral slope is very different at $\ga100$ keV between leptonic and hadronic models. In Fig.\ref{Fig:M2} we show the broadband (radio to TeV) fitting of leptonic model using Naima, where the radio-to-X-ray emission is dominated by synchrotron radiation of primary electrons and the GeV-TeV emission is dominated by IC radiation of primary electrons. In this model the postshock magnetic field is well constrained to be in the order of $\sim10\mu$G \citep[e.g.,][]{yua11}. Around few hundred keV, the emission is dominated by the IC radiation with a hard spectral slope. However, the spectrum around few hundred keV in the hadronic model will be very soft. For comparison, the SEP radiation assuming the hadronic origin for the GeV-TeV emission is also shown in Fig.\ref{Fig:M2}. The obvious difference between leptonic and hadronic models at $\ga100$ keV is important and needs detailed observation in this region by future sensitive hard X-ray telescopes, with flux sensitivity as good as $\sim10^{-13}\rm erg\,cm^{-2}s^{-1}$. This is still challenging to the upcoming telescopes in the near future (Fig.\ref{Fig:M2}).

\subsection{IC 443 and W44}
\begin{figure}
\vskip -0.0 true cm
\centering
\includegraphics[scale=0.4]{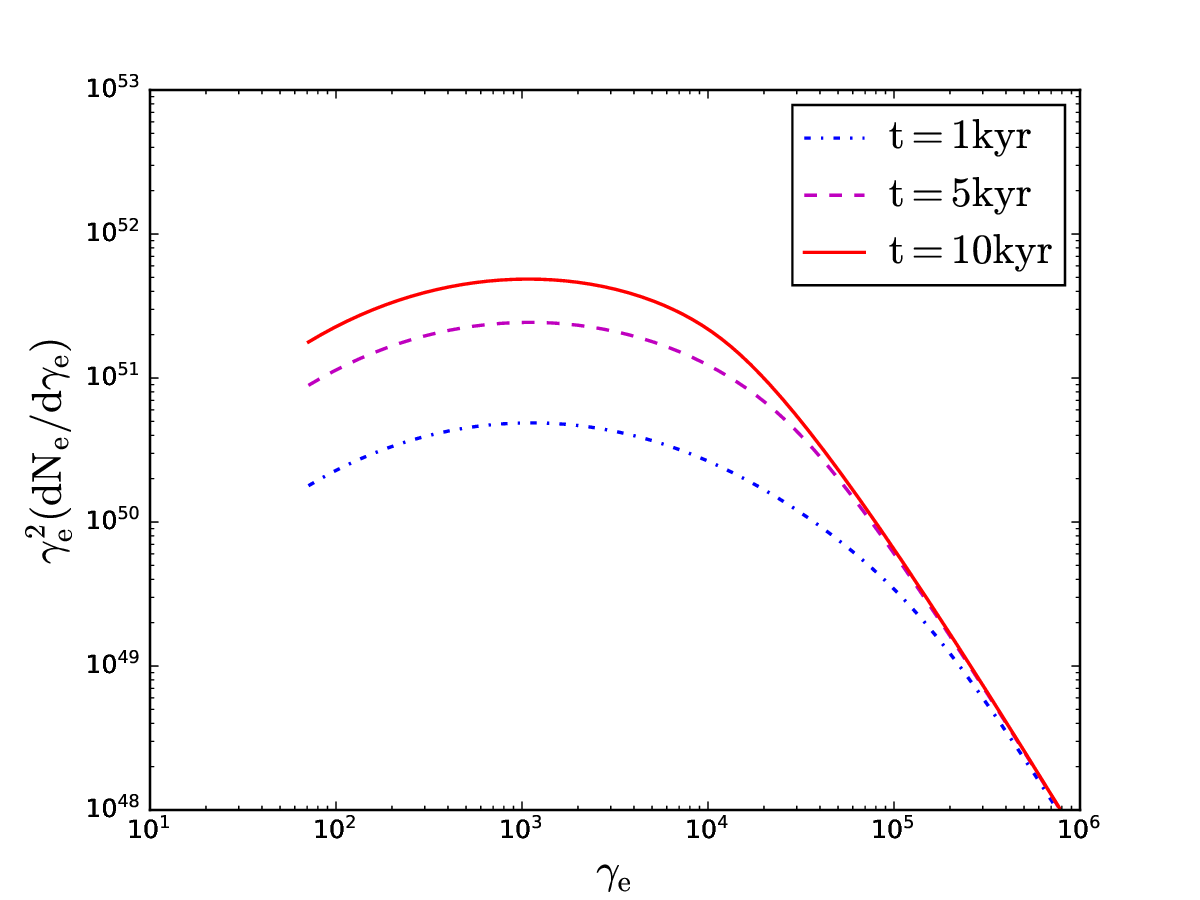}
\includegraphics[scale=0.4]{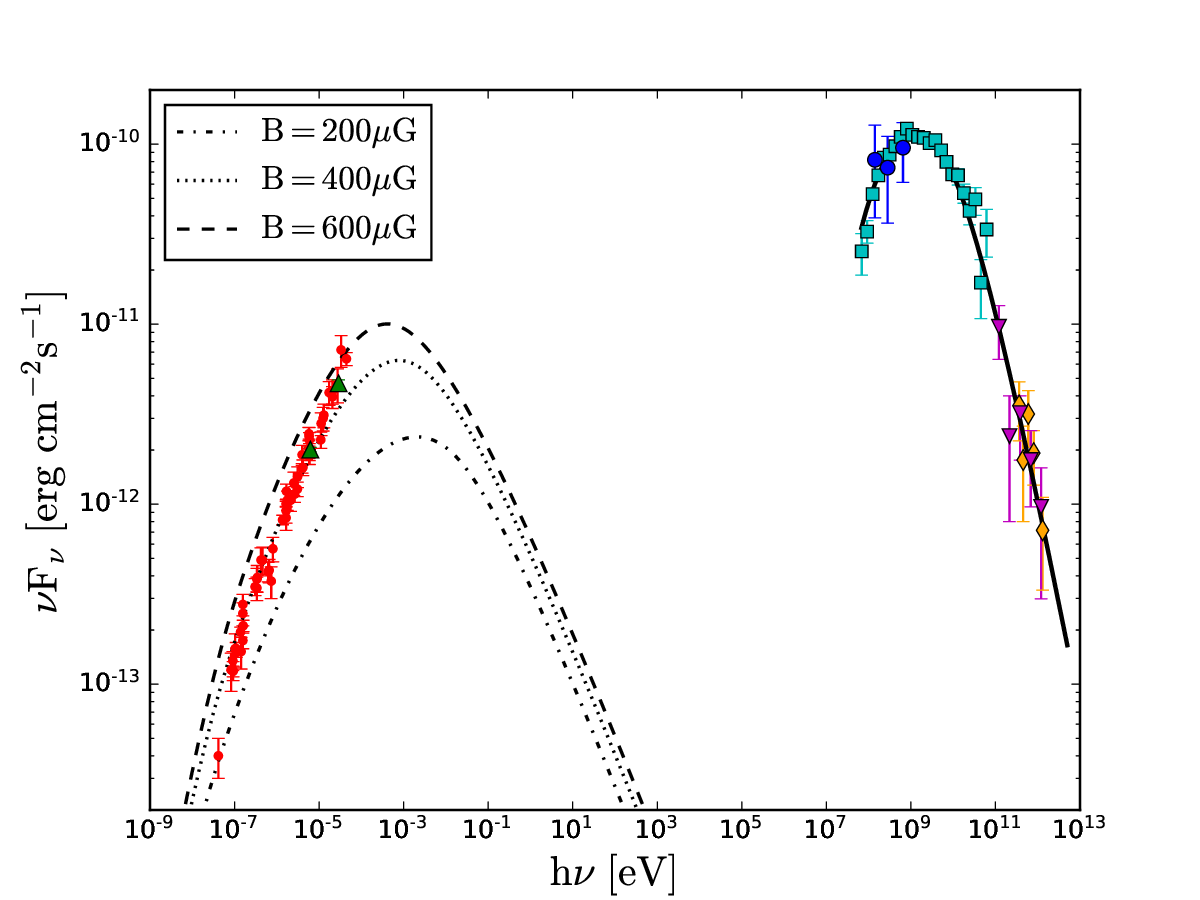}

\caption{Same as Fig.\ref{Fig:M1} but for middle-aged SNR IC 443. Upper panel: the blue dashed-dotted, purple dashed, and red solid lines correspond to $t=1$, 5, and 10 kyr, respectively, with $B=\rm 400 \mu G$. Lower panel: the dashed, dotted, and dashed-dotted lines correspond to $B=600$, 400, and 200 $\rm \mu G$, respectively, at $t=\rm 10 kyr$, and the solid line represents the Eq.(\ref{scpl}) spectral fitting model. Also shown are observational data in radio by VLA (red points) \citep{cas11}, by Sardinia Radio Telescope (green triangles) \citep{egr17}, in $>0.1$ GeV gamma-rays by Fermi-LAT (cyan squares) \citep{ack13}, and by AGILE (blue circulars) \citep{tav10}, and in $>0.1$ TeV gamma-rays by MAGIC (magenta inverted-triangles) \citep{alb07}, and by VERITAS (orange diamonds) \citep{acc09}.}
\label{Fig:IC443}
\end{figure}
\begin{figure}
\vskip -0.0 true cm
\centering
\includegraphics[scale=0.4]{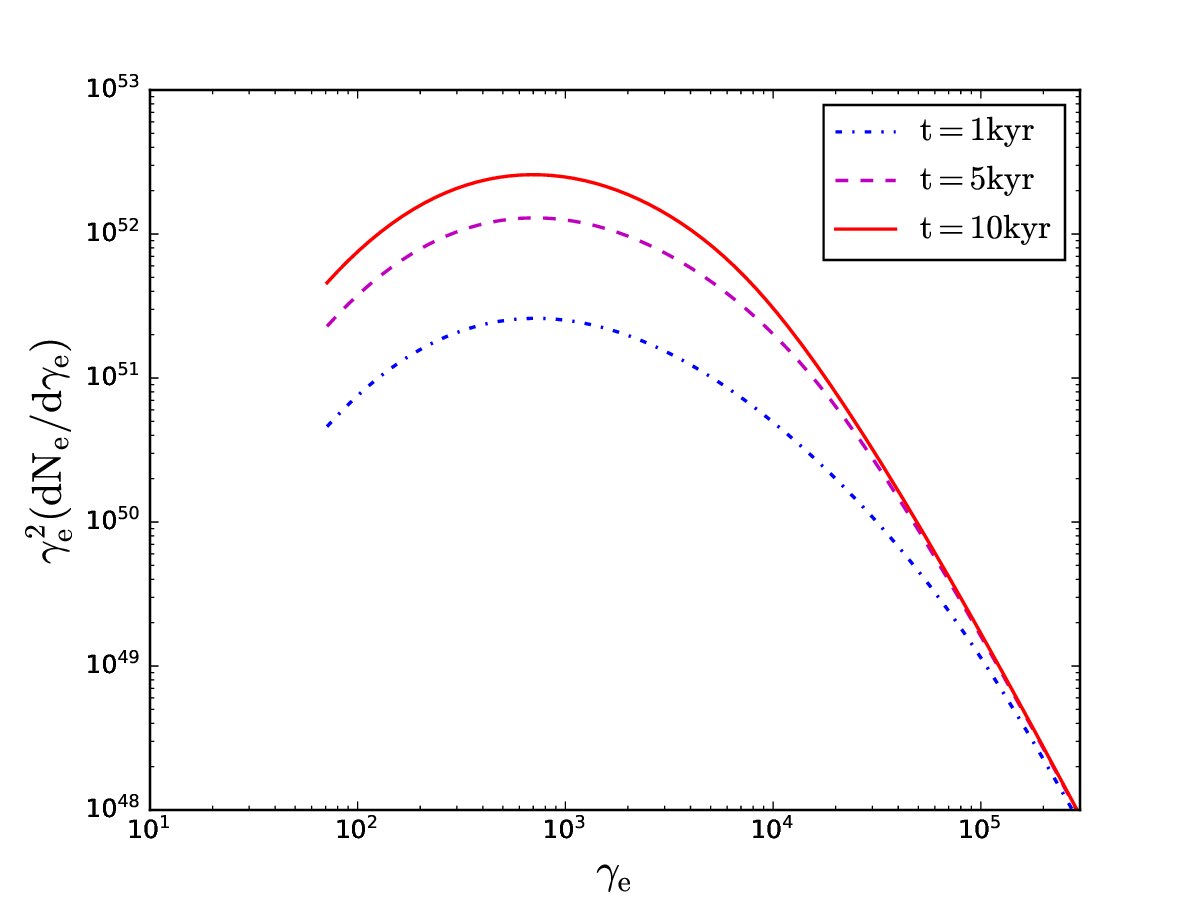}
\includegraphics[scale=0.4]{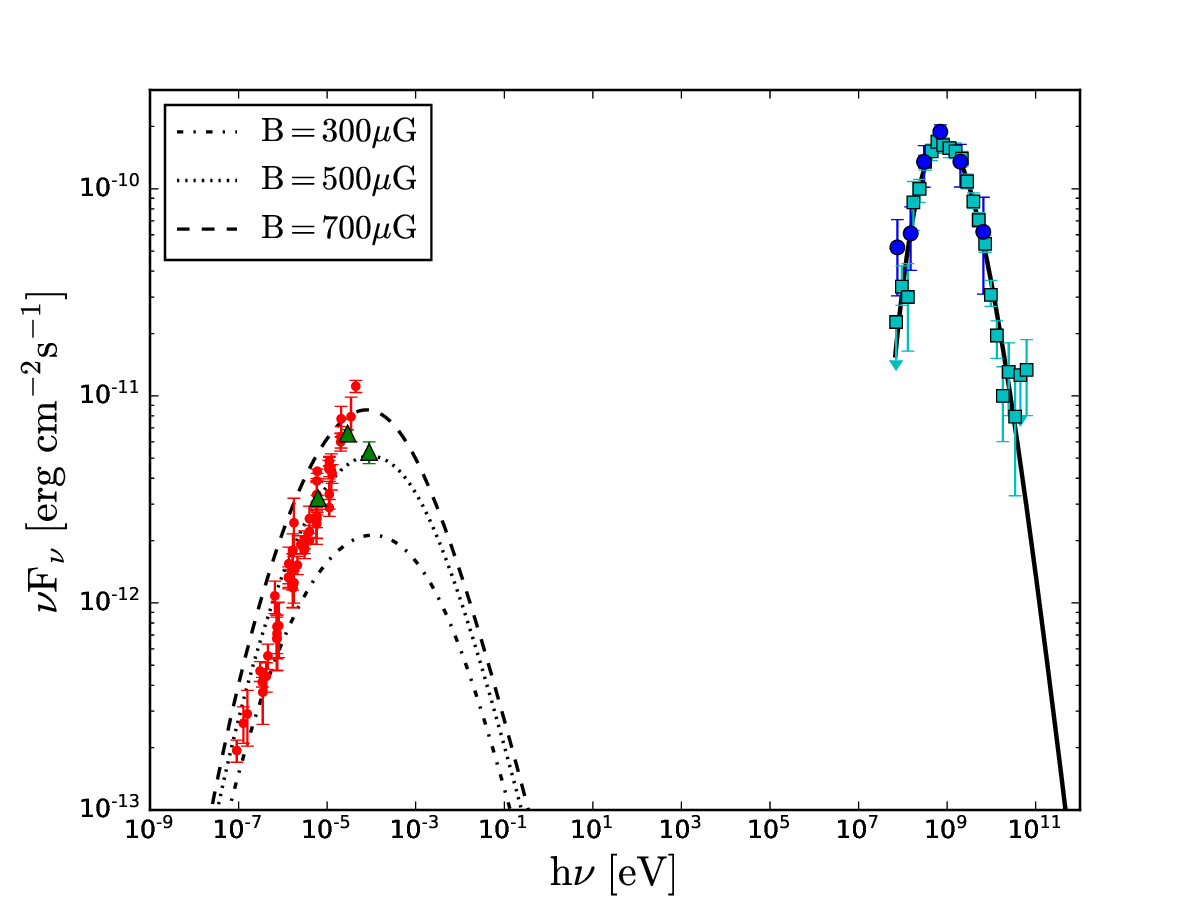}

\caption{Same as Fig.\ref{Fig:M1} but for middle-aged SNR W44. Upper panel: the blue dashed-dotted, purple dashed, and red solid lines correspond to $t=1$, 5, and 10 kyr, respectively, with $B=\rm 500 \mu G$. Lower panel: the dashed, dotted, and dashed-dotted lines correspond to $B=700$, 500, and 300$\rm \mu G$, respectively, at $t=\rm 10 kyr$, and the solid line represents the Eq.(\ref{scpl}) spectral fitting model. Also shown are observational data in radio by VLA (red points) \citep{cas07}, and by Sardinia Radio Telescope (green triangles) \citep{lor18, egr17}, in $>0.1$ GeV gamma-rays by Fermi-LAT (cyan squares) \citep{ack13}, and by AGILE (blue circulars) \citep{giu11}.}
\label{Fig:W44}
\end{figure}

SNR IC 443  and W44 are two middle-aged ($\sim \rm 10kyr$) SNRs surrounded with dense molecular clouds (MCs) \citep{ack13, car14, tav10, fan13,uch10}. Since they are close (with distances of 1.5 kpc and 2.9 kpc for IC 443 and W44, respectively) and bright in gamma-ray energies, they are the best studied SNRs so far. \citet{ack13} shows that the 4-year observations with Fermi-LAT on IC 443 and W44 reveal the characteristic ``pion-decay bump" in the sub-GeV gamma-ray spectra, which tends to exclude a leptonic-only model for the GeV gamma-ray emission from IC 443 and W44. This provides an evidence of CR acceleration in SNRs.

If the GeV emission is produced by hadronic interactions, there should be accompanying SEPs which produce synchrotron radiation in the postshock magnetic field. Again we use the spectral model of Eq.(\ref{scpl}) to fit the gamma-ray data of SNRs IC 443 and W44, and then the best fitting parameters are shown in Tab.\ref{Tab:spectral fitting}. Note, $\xi<0$ indicates that the data can be best fit with a low-energy spectral cutoff. The results of derived SEPs' and synchrotron radiation spectra are shown in Figs.\ref{Fig:IC443} and \ref{Fig:W44} for IC 443 and W44, respectively. We have neglected the attenuation in radio emission due to synchrotron and free-free absorption in the calculation. Indeed the whole radio spectra consistent with a single power law imply negligible absorption in the observational frequency ranges \citep{cas07,cas11}.

We see that the synchrotron flux increases with the magnetic field. Since the predicted flux cannot exceed the observed radio emission, we can give upper limits to the magnetic field, $B\la400\mu$G and $\la500\mu$G for IC 443 and W44, respectively. However, the predicted flux and spectral index are both well consistent with observations, which may implies that the radio emission is mainly produced by the synchrotron radiation of the SEPs. In fact, the observed radio spectral index, if explained by synchrotron radiation, needs a power-law distributed electron population with a spectral index $p=1.75$, which is harder than DSA predicted index $p\approx2$. However, the sub-GeV SEPs, which account for the radio emission, is derived to be hard (as shown in Figs.\ref{Fig:IC443} and \ref{Fig:W44}) because the corresponding gamma-ray spectra at $\la1$~GeV are hard. This can also be seen in Fig 3 of \citet{ack13}, which shows that the proton energy distributions to fit the gamma-ray spectra are hard at $\la10$~GeV. All these facts suggest that the radio emission from IC 443 and W44 is more naturally explained by SEP radiation with magnetic fields $B\simeq400\mu$G and $\simeq 500\mu$G, respectively.

\section{Discussion}
In the hadronic model for RX J1713.7-3946, a spectral bump in $10^{-5}-10^{-2}$eV may show up if the magnetic field is the order of mG (Fig.\ref{Fig:M1}). Observations in microwave to far-infrared bands are encouraged to confirm if it is there, and test the origin of the gamma-rays. However, an upper limit $B\la2$~mG is constrained in this model by the observed radio flux.  \citet {hua08} have obtained comparable result of the magnetic field by carrying out a detailed Monte Carlo particle collision calculation of the SEP production. We here directly use the observed gamma-ray spectrum to derive SEP spectrum analytically, avoiding the complicated numerical calculation of the hadronic interactions.

The synchrotron spectrum from radio to X-ray bands from SNRs can be used to constrain the magnetic field $B$ in the SNR shocks, as discussed in \cite{wangli14}. In the spectrum there is a spectral break $\nu_{\rm cool}$ corresponding to the electrons with a radiative cooling time equal to the SNR age, $h\nu_{\rm cool}\approx3(B/100\unit{\mu G})^{-3}(t/1\unit{kyr})^{-2}$eV, only function of $B$ and $t$ \citep[e.g.,][]{kw08}. If the injected electrons' index is $p$, then the synchrotron spectral slope changes from $F_\nu\propto\nu^{-(p-1)/2}$ below $\nu_{\rm cool}$ to $F_\nu\propto\nu^{-p/2}$ above $\nu_{\rm cool}$. Given the observed radio flux (below $\nu_{\rm cool}$), and the X-ray flux (maybe above $\nu_{\rm cool}$), we can constrain $\nu_{\rm cool}$ and then $B$. For $p\sim1.8$ \cite{hes16} derives $B\sim70\mu$G in RX J1713.7-3946. If $p\ga2$ as expected by DSA theory, $\nu_{\rm cool}$ is larger and $B$ becomes smaller, $\sim10$'s $\mu$G. Small $B$ will lead to enhancement of the inverse-Compton (IC) radiation, and favor leptonic model for the gamma-rays from RX J1713.7-3946.

%If the bulk of gamma-rays from RX J1713.7-3946 is mainly produced by leptonic process, the possible hadronic component should be below the observed flux. A promising flux of pionic gamma-rays is assumed in Fig.\ref{Fig:M2}. The corresponding SEP radiation may show up at $\ga100$keV. For synchrotron radiation at 100 keV, the corresponding pionic gamma-ray energy should be $\sim1$~PeV provided the magnetic field strength $B \sim \rm 10 \mu G$ \citep{wan16}. So a detection of hard X-rays at $100$ keV may provide the evidence of existence of PeV gamma-rays and hence CRs beyond the spectral knee from the SNR. However, a deep observation in the level of $\sim10^{-13}\rm erg\,cm^{-2}s^{-1}$ is required for the future hard X-ray telescopes.

As for IC 443 and W44, the derived SEP radiation can well match the radio data with $B=400\mu$G and $500\mu$G, respectively (Figs \ref{Fig:IC443} and \ref{Fig:W44}). Besides the ``pion-decay bump", these results further support the hadronic origin of gamma-rays. The required relatively large magnetic field, $B\sim100$'s $\mu$G, is consistent with the hadronic model but disfavors leptonic one \citep[e.g.,][]{car14}. Moreover, the large $B$ can be well explained by the large density of the surrounding MCs, since the postshock magnetic field is $B\sim\sqrt{8\pi\epsilon_Bnm_pu_s^2}$ (with $\epsilon_B$ being the energy fraction of the postshock internal energy carried by magnetic field, and $u_{s}$ is the velocity of postshock). For example, for W44 the CO data obtained from the NANTEN2 telescope imply a dense medium with $n \sim 250 -300 \rm cm^{-3}$ \citep{car14}, then the postshock magnetic field is $B\sim0.4(\epsilon_B/10^{-1})^{1/2}(n/200\unit{cm^{-3}})^{1/2}(u_s/300\unit{km\,s^{-1}})$mG\footnote{ See, however, that in some recent simulations, the non-relativistic or mildly relativistic shocks typically have much lower values of $\epsilon_B$, i.e., $\epsilon_B < 10^{-3}$ \citep{cap14, van18, cru19}, which is in tension with the hadronic models of high energy emission from SNRs.}, consistent with the requirement for SEP radiation to account for the radio emission.

Note, \citet{uch10} had shown that, in a specific model where the medium is clumpy clouds, the SEP synchrotron radiation can dominate primary electron radiation and account for the observed radio spectrum from SNR W44. In their model the SEPs and secondary gamma-rays are produced in the same region (i.e., the shocked clumpy clouds), a situation similar to our one-zone case, thus we can make comparison between their and our results. It can be found following their calculation that the required magnetic field of shocked clouds in their specific model is $\sim460\rm \mu G$, similar to our result derived directly from gamma-ray data. This also shows the advantage of our method that no CR spectrum and $pp$ interaction should be considered in the calculation, but simply using the correlation between the SEP and secondary gamma-ray productions.

The measurement of the postshock magnetic field is important for the study of the collisionless shock physics. In any case, by requiring the SEP synchrotron radiation not exceeding the observed radio-to-X-ray flux we can give an upper limit of the magnetic field. This is simply true if the gamma-rays are hadronic origin. If, however, the gamma-rays are leptonic origin, an even low magnetic field is required to model the radio to gamma-ray data (This is because in the leptonic models the radio to X-ray emission is explained by synchrotron radiation, whereas the gamma-ray emission by IC or bremsstrahlung. A low magnetic field can suppress the synchrotron radiation so that the IC and or bremsstrahlung radiation can be enhanced). So after all, the requirement of SEP synchrotron radiation below observed radio flux always gives an upper limit of the magnetic field. Thus we can conclude that $B\la$ few mG in the shock of RX J1713.7-3946, and $B\la0.5$mG in the shocks of IC 443 and W44.

We can estimate the efficiency that primary electrons can be accelerated in IC 443 and W44, based on the conclusion that the entire spectrum is contributed by the hadronic gamma-rays and the SEP synchrotron radiation, rather than primary electrons. The proton lifetime is $t_{\rm pp}=6 \times 10^7 (n/1 \rm cm^{-3})^{-1} yr$ \citep{aha96}. With $n \sim 200 \rm cm^{-3}$ and $t_{\rm SNR} \sim 10 \rm kyr$, the $pp$ pion production efficiency is estimated to be $t_{\rm SNR}/t_{pp} \sim 3\times10^{-2}$. A fraction $1/6$ of proton energy goes to the SEP energy, so the total energy of the SEPs is a fraction $5\times10^{-3}$ of the primary protons. Since the total energy of the primary electrons should be lower than that of the SEPs, the primary electron to proton ratio should be lower than $5\times10^{-3}$. This should be compared to the electron-proton ratio, $\gtrsim 10^{-3}$, implied by the radio observations of SNRs in nearby galaxies \citep{kw08, zhang16}.

\section{Conclusion and summary}
Since there is a connection between the SEPs and gamma ray photons, there should be a connection between the synchrotron radiation produced by SEPs and the observed gamma-rays from hadronic interactions. Here, we develop a simple analytical  method to transform the observed gamma-ray spectra to the SEP spectra, and then calculate the SEP synchrotron radiation. We apply our method to three well observed gamma-ray SNRs, RX J1713.7-3946, IC 443 and W44. The main results are:
\begin{itemize}
\item If the GeV-TeV gamma-rays from young SNR RX J1713.7-3946 are produced by hadronic interactions, the SEP synchrotron radiation may give rise to a spectral bump at $10^{-5}-10^{-2}$eV dominating the predicted synchrotron component in leptonic model, if the magnetic field in the SNR shock is $\ga0.5$mG. In hard X-ray range, $\ga100$ keV, the SEP synchrotron radiation may show up as an excess compared to the synchrotron spectrum from primary electron contribution, which is a distinct feature of hadronic model but needs future sensitive hard X-ray telescopes to observe.
%\item If the GeV-TeV gamma-rays from young SNR RX J1713.7-3946 are dominated by the contribution from leptonic processes, there may be a subdominant hadronic gamma-ray component, which can be accompanied by a SEP synchrotron radiation that may be detected at $\ga100$ keV, with a flat spectrum and a flux of a few $10^{-13}\unit{erg\,cm^{-2}s^{-1}}$.
\item If the GeV-TeV gamma-rays from middle-aged SNRs IC 443 and W44 are produced by hadronic interactions, the SEP synchrotron radiation can well account for the observed radio flux and spectral slopes, which further supports the hadronic origin of the gamma-rays and the TeV CR accelerations in these two SNRs.
\item The requirement that the SEP radiation derived by the observed gamma-ray spectra, assumed to be hadronic origin, cannot exceed the observed radio flux constrain the magnetic fields in the SNR shocks to be $B\la 2$mG, $\la 400 \mu$G, and $\la500 \mu$G for RX J1713.7-3946, IC 443 and W44, respectively.
\end{itemize}

In brief, the SEP synchrotron radiation can be a powerful tool to distinguish production processes of the GeV-TeV gamma-ray emission from SNRs. Future microwave to far infrared and deep hard X-ray observations may tell the differences between the synchrotron spectra by SEPs and primary electrons, and probe the particle accleration ability of SNR shocks.

\section*{Acknowledgements}
We thank the anonymous referees for valuable suggestions. We also thank Xiao Zhang and Ke Wang for helpful discussion. This work is partly supported by the Natural Science Foundation of China (No. 11773003, No. 11622326, No. 11203067, No. 11721303, and No. U1931201), the 973 Program of China (No. 2014CB845800), the National Program on Key Research and Development Project (Grants No. 2016YFA0400803) and the Yunnan Natural Science Foundation (No. 2011FB115 and No. 2014FB188).

%%%%%%%%%%%%%%%%%%%%%%%%%%%%%%%%%%%%%%%%%%%%%%%%%%



%\bibliographystyle{mnras}


%%%%%%%%%%%%%%%%%%%%%%%%%%%%%%%%%%%%%%%%%%%%%%%%%%

% Don't change these lines
\bsp	% typesetting comment
\label{lastpage}
\end{document}